\documentstyle[preprint,aps,epsf,eqsecnum]{revtex}
\tightenlines
\begin{document}
\def\non{\nonumber}
\def\be{\begin{eqnarray}}
\def\en{\end{eqnarray}}
\def\la{\langle}
\def\ra{\rangle}
\def\pr{{\sl Phys. Rev.}~}
\def\prl{{\sl Phys. Rev. Lett.}~}
\def\pl{{\sl Phys. Lett.}~}
\def\np{{\sl Nucl. Phys.}~}
\def\zp{{\sl Z. Phys.}~}
\draft
\vfill
\title{\Large\bf
A consistent treatment for pion form factors \\
in space-like and time-like regions
}

\vfill
\author{Chien-Wen Hwang}
\address{\rm Department of Physics, National Tsing Hua University, Hsinchu 300, Taiwan \rm}
\vfill
\maketitle
%
\begin{abstract}
We write down some relevant matrix elements for the scattering and decay processes of the pion by considering a quark-meson vertex function. The pion charge and transition form factors $F_\pi$, $F_{\pi\gamma}$, and $F_{\pi\gamma^*}$ are extracted from these matrix elements using a relativistic quark model on the light-front. We found that, the form factors $F_\pi$ and $F_{\pi\gamma}$ in the space-like region agree well with experiment. Furthermore, the branching ratios of all observed decay modes of the neutral pion, that are related to the form factors $F_{\pi\gamma}$ and $F_{\pi\gamma^*}$ in the time-like region, are all consistent with the data as well. Additionally, $F_\pi$ in the time-like region, which deals with the nonvalence contribution, is also discussed.
\end{abstract}
\pagestyle{plain}
\section{Introduction}
Form factors are important physical quantities in the understanding of the internal structure of hadrons and relate to many other physical quantities. In this paper, we study two types of form factors for the pion: charge and transition ones. On the one hand, the former appears in the calculation of elastic electron-pion scattering in which one off-shell photon exchanges between electron and one of the quarks in pion. This form factor is related to the electromagnetic radius of the charged pion. The latter, on the other hand, comes from the reactions where the pion is coupled to two photons. This form factor also interrelates to the decay rates and branching ratios of all the observed decay modes of neutral pion. It is well known that these form factors must be treated with non-perturbative method. There are many different candidates for this purpose, such as lattice calculations \cite{Dan}, vector meson dominance (VMD) \cite{GL,ABBM}, perturbative QCD (pQCD) with some non-perturbative input parameters \cite{LB,LS,KO,ADT}, QCD sum rules \cite{SVZ,BH,Kho}, and the light-front quark model (LFQM) \cite{CCC,Card,CJ2}.

LFQM is the only relativistic quark model in which a consistent and fully relativistic treatment of quark spins and the center-of-mass motion can be carried out. Thus it has been applied in the past to calculate various form factors \cite{CCC,Card,CJ2,Jaus,Don,Dem,CHZ,CCH}. This model has many advantages. For example, the light-front wave function is manifestly invariant under boost as it is expressed in terms of the momentum fraction variables (in ``+" component) in an analogy to the parton distributions in the infinite momentum frame. Moreover, hadron spin can also be relativistically constructed by using the so-called Melosh rotation. The kinematic subgroup of the light-front formalism has the maximum number of interaction-free generators including the boost operator which describes the center-of-mass motion of the bound state (for a review of the light-front dynamics and light-front QCD, see \cite{Zhang}). On the one hand, we will concentrate on the space-like region $q^2\leq 0$ ($q$ being the momentum transfer) for charge form factors. In this region, the so-called Z graph contribution \cite {CCH} vanishes and only the valence-quark contributes. As for the time-like region $q^2\geq 0$ for charge form factor, it concerns with the nonvalence contribution (Z graph). A reliable way of estimating this part is still lacking. On the other hand, the space-like and time-like regions, which correspond to the scattering and decay processes, respectively, are both considered for transition form factors. We make a consistent treatment with LFQM for the decay constants, the charge form factors, and the transition form factors which include the Z graph contribution within the time-like region. It must be emphasized that these derivations could be applied to all of the kinematically allowed region, no matter the momentum transfer is large or small. We compare these results with some experimental data on the charge form factors \cite{Amen,Bebek,Volmer,Brauel}, transition form factors \cite {CLEOFpgs}, and branching ratios of some decay modes \cite{PDG00} for the pion.

The paper is organized as follows. In Sec. II, the basic theoretical formalism is given and the decay constant, the charge and transition form factors are derived for the pion. In Sec. III,  we fix the parameters appearing in the wave functions and calculate the form factors and branching ratios. Finally, a conclusion is given in Sec. IV.

\section{Framework}
In this section, we first write down some relevant matrix elements in the covariant form. After integrating out the minus component of the internal monentum, we introduce the light-front quark model to formulate the decay constant, charge form factors and transition form factors.
\subsection{decay constant}
The decay constant of a pion is defined by
\be
\la 0|A^\mu|\pi(p)\ra=\,\sqrt{2}i~f_\pi~ p^\mu,\label{decaydef}
\en
where $A^\mu=\bar q_2 \gamma^\mu \gamma_5 q_1$ is the axial vector current. Assuming a constant vertex function $\Lambda_\pi$ \cite{Jaus,Dem} which is related to the $q\bar q$ bound state of the pion.
Then the quark-meson diagram, depicted in Fig.1 (a), yields
\be
\la 0|\bar q_2 \gamma^\mu \gamma_5 q_1|\pi(p)\ra=-\sqrt{N_c}\Lambda_\pi\int {d^4p_1\over{(2\pi)^4}} \text{Tr}\Bigg[\gamma_5{i(\not{\!p_1}-\not{\!p}+m)\over{(p_1-p)^2-m^2+i\epsilon}}\gamma^\mu\gamma_5{i(\not{\!p_1}+m)\over{p_1^2-m^2+i\epsilon}}\Bigg], \label{decay1}
\en
where $m=m_u=m_d$ and $N_c$ is the number of colors. In terms of the LF coordinates $(p_1^-,p_1^+,{p_1}_\perp)$, we obtain
\be
\la 0|\bar q_2 \gamma^\mu \gamma_5 q_1|\pi(p)\ra &=&-{\sqrt{N_c}\Lambda_\pi\over{2(2\pi)^4}}(4 p^\mu)\int {dp_1^-dp_1^+d^2{p_1}_\perp\over{p_1^+(p_1-p)^+}}m\Bigg[\Bigg(p^-_1-{m^2+{p_1^2}_\perp\over{p_1^+}}+{i\epsilon\over{p_1^+}}\Bigg) \non \\ &&\times\Bigg(p^-_1-p^--{m^2+(p_1-p)^2_\perp\over{(p_1^+-p^+)}}+{i\epsilon\over{(p_1^+-p^+)}}\Bigg)\Bigg]^{-1}.  \label{decay2}
\en
Comparing (\ref{decaydef}) with (\ref{decay2}) and performing the integration over the LF ``energy" $p_1^-$ in (\ref{decay2}), we obtain
\be
f_\pi&=&2\sqrt{2}\sqrt{N_c}\int{dp_1^+d^2{p_1}_\perp\over{2(2\pi)^3}} {m\over{p_1^+(p^+-p^+_1)}}\Bigg[\Lambda_\pi\Bigg(p^--{m^2+(p_1-p)^2_\perp\over{p^+-p^+_1}}-{m^2+{p_1^2}_\perp\over{p_1^+}}\Bigg)^{-1}\Bigg]. \label{decay4}
\en
\subsection{charge form factor}
The charge form factor of the pion in the space-like region, as being illustrated in Fig.1 (b), is determined by the scattering of one virtual photon and one pion. This form factor can be defined by the matrix element
\be
\la \pi'(p')|J^\mu|\pi(p)\ra = e~F^{\text{s}}_\pi(q^2) (p+p')^\mu, \label{FPdef}
\en
where $J^\mu = \bar q e_q e\gamma^\mu q$ is the vector current, $e_q$ is the charge of quark $q$ in unit of $e$, $q^2=(p'-p)^2\leq 0$, and the superscript ``s" represents the space-like region. From Fig.1 (b), we obtain
\be
\la \pi'(p')|\bar q e_q e \gamma^\mu q|\pi(p)\ra &=&-(e_u+e_{\bar d}) e\Lambda_\pi\Lambda_{\pi'}\int {d^4p_1\over{(2\pi)^4}} \text{Tr}\Bigg[\gamma_5{i(-\not{\!p_1}+m)\over{p_1^2-m^2+i\epsilon}}\gamma_5\non \\
&&~~~~\times{i(\not{\!p'}-\not{\!p_1}+m)\over{(p'-p_1)^2-m^2+i\epsilon}}e\gamma^\mu{i(\not{\!p}-\not{\!p_1}+m)\over{(p-p_1)^2-m^2+i\epsilon}}\Bigg], \label{charge1}
\en
Using the LF coordinates and taking the trace, we obtain
\be
\la \pi'(p')|\bar q e_q e \gamma^\mu q|\pi(p)\ra &=&-{e\Lambda_\pi\Lambda_{\pi'}\over{2(2\pi)^4}}\int {dp_1^-dp_1^+d^2{p_1}_\perp\over{p_1^+(p-p_1)^+(p'-p_1)^+}} \Bigg\{I^\mu\Bigg[\Bigg(p^-_1-{m^2+{p_1^2}_\perp\over{p_1^+}}+{i\epsilon\over{p_1^+}}\Bigg)\non \\
&&\times\Bigg(p^--p^-_1-{m^2+(p-p_1)^2_\perp\over{p^+-p^+_1}}+{i\epsilon\over{p^+-p^+_1}}\Bigg)\non \\
&&\times\Bigg(p'^--p^-_1-{m^2+(p'-p_1)^2_\perp\over{p'^+-p^+_1}}+{i\epsilon\over{p'^+-p_1^+}}\Bigg)\Bigg]^{-1}\Bigg\}, \label{charge2}
\en
where
\be
I^\mu&=&4\Bigg[p_1^\mu(p_1^2-p\cdot p'-m^2)+p^\mu(-p_1^2+p_1\cdot p'+m^2)+p'^\mu(-p_1^2+p_1\cdot p+m^2)\Bigg].
\en
Let us extract $F_\pi(Q^2)$ from the ``+" component of the vector current $J^\mu$. It does not loss the generality if $q^+$ is set to be zero for the momentum transfer in the space-like region. Performing the $p_1^-$-integration, we obtain
\be
F^{\text{s}}_\pi(q^2)&=&-\int {dp_1^+d^2{p_1}_\perp\over{2(2\pi)^3}}\Bigg\{{(e_u+e_{\bar d}) \tilde{I}\over{p_1^+(p-p_1)^+(p'-p_1)^+}} \Bigg[\Lambda_\pi\Bigg(p^--{m^2+(p-p_1)^2_\perp\over{p^+-p^+_1}}-{m^2+{p_1^2}_\perp\over{p_1^+}}\Bigg)^{-1}\Bigg]\non \\
&&~~~~~~~~~~~~~~\times\Bigg[\Lambda_{\pi'}\Bigg(p'^--{m^2+(p'-p_1)^2_\perp\over{p'^+-p^+_1}}-{m^2+p_{1\perp}^2\over{p_1^+}}\Bigg)^{-1}\Bigg]\Bigg\},
\en
where
\be
\tilde{I} \equiv {I^+\over{2p^+}}\Bigg|_{q^+=0,~p_1^-={m^2+p_{1\perp}^2\over{p^+_1}}}
\en

By contrast, when we consider the process $\gamma^*\to \pi\pi$ which is illustrated in Fig. 1(c), the momentum transfer $q^2\geq 4M^2_\pi$ is in the time-like region. 
In this region, the charge form factor is determined by another matrix element
\be
\la \pi'(p')\pi(p)|\bar q e \gamma^\mu q|0 \ra &=&-(e_u+e_{\bar d}) e\Lambda_\pi\Lambda_{\pi'}\int {d^4p_1\over{(2\pi)^4}} \text{Tr}\Bigg[\gamma_{\mu}{i(-\not{\!p_1}+m)\over{p_1^2-m^2+i\epsilon}}\gamma_5\non \\
&&~~~~\times{i(\not{\!p}-\not{\!p_1}+m)\over{(p-p_1)^2-m^2+i\epsilon}}e\gamma^\mu{i(\not{\!q}-\not{\!p_1}+m)\over{(q-p_1)^2-m^2+i\epsilon}}\Bigg].\label{timeME}
\en
where $q=(p+p')$.
It is well known that $q^2=q^+q^--q^2_\perp$ in the LF coordinate, if we firstly assume \cite{DK} the momentum transfer is purely longitudinal, i.e., $q_\perp=0$, $q^2=q^+q^-\geq 0$ will be ensured. The same results will be obtained if one carries out the integral and then takes $q_\perp=0$. However, the former method will reduce the calculating processes considerably. Thus this form factor is obtained as
\be
F^{\text{t}}_\pi(q^2)&=&-\int {d^2{p_1}_\perp\over{2(2\pi)^3}}\int^{p^+}_0 dp_1^+{\tilde{K_1}\Lambda_{\pi'}\over{p_1^+(q-p_1)^+(p-p_1)^+}} \Bigg(q^--{m^2+(q-p_1)^2_\perp\over{q^+-p^+_1}}-{m^2+{p_1^2}_\perp\over{p_1^+}}\Bigg)^{-1}\non \\
&&~~~~~~~~~~~~~~\times\Bigg[\Lambda_\pi\Bigg(p^--{m^2+(p-p_1)^2_\perp\over{p^+-p^+_1}}-{m^2+p_{1\perp}^2\over{p_1^+}}\Bigg)^{-1}\Bigg] \non \\
&-&\int {d^2{p_1}_\perp\over{2(2\pi)^3}}\int^{^q+}_{p^+}dp_1^+{\tilde{K_2}\Lambda_{\pi}\over{p_1^+(q-p_1)^+(p_1-p)^+}} \Bigg(q^--{m^2+(q-p_1)^2_\perp\over{q^+-p^+_1}}-{m^2+{p_1^2}_\perp\over{p_1^+}}\Bigg)^{-1}\non \\
&&~~~~~~~~~~~~~~\times\Bigg[\Lambda_{\pi'}\Bigg(p'^--{m^2+(p-p_1)^2_\perp\over{p^+_1-p^+}}-{m^2+p_{1\perp}^2\over{(q-p_1)^+}}\Bigg)^{-1}\Bigg]
\en
where the superscript ``t" represents the time-like region,
\be
\tilde{K_1}={K^+\over{2q^+}}\Bigg|_{p_1^-={m^2+p_{1\perp}^2\over{p^+_1}}},~~~\tilde{K_2}={K^+\over{2q^+}}\Bigg|_{(q-p_1)^-={m^2+(q-p_1)_\perp^2\over{(q-p_1)^+}}},
\en
and
\be
K^\mu&=&4\Bigg[p_1^\mu(m^2-p_1^2+2p_1\cdot p-p\cdot q)+p^\mu(-p_1^2+p_1\cdot q+m^2)+q^\mu(p_1^2-p_1\cdot p-m^2)\Bigg].
\en
\subsection{transition form factor}
There are two types of transition form factors, $F_{\pi\gamma}$ and $F_{\pi\gamma^*}$. The former comes from a neutral pion which coupled to two photons with one on-shell and the other off-shell, while the latter arises from  the $\pi^0\gamma^*\gamma^*$ vertex, where $\gamma^*$ represent an off-shell photons. These two form factors are related to many interesting experimental data. In this section, we will calculate them in detail.

\subsubsection{transition form factor $F_{\pi\gamma}$}
The form factor $F_{\pi\gamma}$ is defined by the $\pi\gamma\gamma^*$ vertex\cite{LB}
\be
\Gamma_\mu=-ie^2~F_{\pi\gamma} (q^2)\epsilon_{\mu\nu\rho\sigma} p^\nu q^\rho \varepsilon^\sigma, \label{tran1}
\en
where $q$($\varepsilon$) is the momentum (polarization) of the on-shell photon. This process is illustrated in Fig.1 (e). The amplitude is given by
\be
\Gamma_\mu&=&-e_q e_{\bar q'}e^2\sqrt{N_c}\Lambda_\pi\int {d^4p_1\over{(2\pi)^4}} \non \\
&&\times\Bigg\{\text{Tr}\Bigg[\gamma_5 {i(\not{\!p_1}-\not{\!p}+m)\over{(p_1-p)^2-m^2+i\epsilon}}\gamma_\mu{i(\not{\!p_1}-\not{\!p}-\not{\!q}+m)\over{(p_1-p-q)^2-m^2+i\epsilon}}\not{\!\varepsilon}{i(\not{\!p_1}+m)\over{p_1^2-m^2+i\epsilon}}\Bigg]\non\\
&&~~+\text{Tr}\Bigg[\gamma_5 {i(\not{\!p_1}-\not{\!p}+m)\over{(p_1-p)^2-m^2+i\epsilon}}\not{\!\varepsilon}{i(\not{\!p_1}+\not{\!q}+m)\over{(p_1+q)^2-m^2+i\epsilon}}\gamma_\mu{i(\not{\!p_1}+m)\over{p_1^2-m^2+i\epsilon}}\Bigg]\Bigg\}. \label{tran2}
\en
Using the light-front coordinates and taking the trace in(\ref{tran2}), we obtain
\be
&&\Gamma_\mu=e_q e_{\bar q'}e^2\sqrt{N_c}\Lambda_\pi\int {dp_1^-dp_1^+d^2{p_1}_\perp\over{2(2\pi)^4}}4im\epsilon_{\mu\nu\rho\sigma} p^\nu q^\rho \varepsilon^\sigma  \non \\
&&~~~~\times\Bigg\{\Bigg[p_1^+(p_1-p)^+(p_1+q)^+\Bigg(p^-_1-p^--{m^2+(p_1-p)^2_\perp\over{p_1^+-p^+}}+{i\epsilon\over{p_1^+-p^+}}\Bigg)\non \\
&&~~~~~~\times\Bigg(p^-_1-{m^2+{p_1^2}_\perp\over{p_1^+}}+{i\epsilon\over{p_1^+}}\Bigg)\Bigg(p^-_1+q^--{m^2+(p_1+q)^2_\perp\over{p_1^++q^+}}+{i\epsilon\over{p_1^++q^+}}\Bigg)\Bigg]^{-1} \non \\
&&~~~~~~+\Bigg[p_1^+(p_1-p)^+(p_1-p-q)^+\Bigg(p^-_1-p^--{m^2+(p_1-p)^2_\perp\over{p_1^+-p^+}}+{i\epsilon\over{p_1^+-p^+}}\Bigg)\non \\
&&~~~~~~\times\Bigg(p^-_1-{m^2+{p_1^2}_\perp\over{p_1^+}}+{i\epsilon\over{p_1^+}}\Bigg)\Bigg(p^-_1-p^--q^--{m^2+(p_1-p-q)^2_\perp\over{p_1^+-p^+-q^+}}+{i\epsilon\over{p_1^+-p^+-q^+}}\Bigg)\Bigg]^{-1}\Bigg\}. \non \\
\label{tran3}
\en
$F_{\pi\gamma}$ can be extracted easily by comparing (\ref{tran1}) with (\ref{tran3}). When we consider $q^2\leq 0$, the $p^-_1$-integration is performed in the condition $q^+=0$ and the result is
\be
F^{\text{s}}_{\pi\gamma}(q^2)&=&-4\sqrt{N_c}e_q e_{\bar q'}\int{dp_1^+d^2{p_1}_\perp\over{2(2\pi)^3}}m\Bigg[\Lambda_\pi\Bigg(p^--{m^2+(p_1-p)^2_\perp\over{p^+-p^+_1}}-{m^2+{p_1^2}_\perp\over{p_1^+}}\Bigg)^{-1}\Bigg]\non \\
&\times&\Bigg\{\Bigg[p_1^{+2}(p-p_1)^+\Bigg((p+q)^--{m^2+(p_1-p)^2_\perp\over{p^+-p^+_1}}-{m^2+(p_1+q)^2_\perp\over{p_1^+}}\Bigg)\Bigg]^{-1} \non \\
&&+\Bigg[p_1^+(p-p_1)^{+2}\Bigg((p+q)^--{m^2+(p_1-p-q)^2_\perp\over{p^+-p^+_1}}-{m^2+(p_1+q)^2_\perp\over{p_1^+}}\Bigg)\Bigg]^{-1}\Bigg\}. \label{tran4}
\en

If the decay process $q^2\geq 0$ is considered, the momentum transfer is in the time-like region. The $p^-_1$-integration is performed in the condition $q_\perp =0$ in two regions: $0\leq p^+_1\leq q^+$ and $q^+\leq p^+_1\leq p^+$, and in their photons-exchanged part: $0\leq p^+_1\leq (p-q)^+$ and $(p-q)^+\leq p^+_1\leq p^+$
\be
F^{\text{t}}_{\pi\gamma}(q^2)&=&4\sqrt{N_c}e_q e_{\bar q'}m \int {d^2{p_1}_\perp\over{2(2\pi)^3}} \Bigg\{\int^{q^+}_0 dp_1^+ \Bigg[\Lambda_\pi\Bigg(p^--{m^2+(p_1-p)^2_\perp\over{p^+-p^+_1}}-{m^2+{p_1^2}_\perp\over{p_1^+}}\Bigg)^{-1}\Bigg]\non \\
&&\times\Bigg[p_1^+(p-p_1)^+(q-p_1)^+\Bigg(q^--{m^2+{p_1^2}_\perp\over{p^+_1}}-{m^2+{p_1^2}_\perp\over{q^+-p_1^+}}\Bigg)\Bigg]^{-1} \non \\
&&+\int^{p^+}_{q^+} dp_1^+ \Bigg[\Lambda_\pi\Bigg(p^--{m^2+(p_1-p)^2_\perp\over{p^+-p^+_1}}-{m^2+{p_1^2}_\perp\over{p_1^+}}\Bigg)^{-1}\Bigg]\non \\
&&\times\Bigg[p_1^+(p-p_1)^+(p_1-q)^+\Bigg((p-q)^--{m^2+{(p_1-p)^2}_\perp\over{p^+-p^+_1}}-{m^2+{p_1^2}_\perp\over{p_1^+-q^+}}\Bigg)\Bigg]^{-1} \non \\
&&+\int^{p^+-q^+}_0 dp_1^+ \Bigg[\Lambda_\pi\Bigg(p^--{m^2+(p_1-p)^2_\perp\over{p^+-p^+_1}}-{m^2+{p_1^2}_\perp\over{p_1^+}}\Bigg)^{-1}\Bigg]\non \\
&&\times\Bigg[p_1^+(p-p_1)^+(p-q-p_1)^+\Bigg((p-q)^--{m^2+{(p-p_1)^2}_\perp\over{(p-p_1)^+}}-{m^2+{(p-p_1)^2}_\perp\over{(p-q-p_1)^+}}\Bigg)\Bigg]^{-1} \non \\
&&+\int_{p^+-q^+}^{p^+} dp_1^+ \Bigg[\Lambda_\pi\Bigg(p^--{m^2+(p_1-p)^2_\perp\over{p^+-p^+_1}}-{m^2+{p_1^2}_\perp\over{p_1^+}}\Bigg)^{-1}\Bigg]\non \\
&&\times\Bigg[p_1^+(p-p_1)^+(p_1-p+q)^+\Bigg(q^--{m^2+(p_1-p)^2_\perp\over{(p_1-p+q)^+}}-{m^2+{(p_1-p)^2}_\perp\over{(p-p_1)^+}}\Bigg)\Bigg]^{-1}\Bigg\}.\non \\
\label{trant3}
\en

\subsubsection{transition form factor $F_{\pi\gamma^*}$}
As for the form factor $F_{\pi\gamma^*}$, it is defined by the $\pi\gamma^*\gamma^*$ vertex \cite{KWZ}
\be
\Gamma_{\mu\nu} = -ie^2F_{\pi\gamma^*}(q^2,q'^2) \varepsilon_{\mu\nu\rho\sigma} {\cal Q}^\rho {\cal P}^\sigma, \label{trana}
\en
where ${\cal Q}\equiv {1\over{2}}(q'-q)$, ${\cal P}\equiv q'+q$, and $q'=p+q$. This process is depicted in Fig.1 (f). The amplitude can be written as
\be
\Gamma_{\mu\nu}&=&-e_q e_{\bar q'}e^2\sqrt{N_c}\Lambda_\pi\int {d^4p_1\over{(2\pi)^4}} \non \\
&&\times\Bigg\{\text{Tr}\Bigg[\gamma_5 {i(\not{\!p_1}-\not{\!p}+m)\over{(p_1-p)^2-m^2+i\epsilon}}\gamma_\mu{i(\not{\!p_1}-\not{\!p}-\not{\!q}+m)\over{(p_1-p-q)^2-m^2+i\epsilon}}\gamma_\nu{i(\not{\!p_1}+m)\over{p_1^2-m^2+i\epsilon}}\Bigg] \non \\
&&~~+\text{Tr}\Bigg[\gamma_5 {i(\not{\!p_1}-\not{\!p}+m)\over{(p_1-p)^2-m^2+i\epsilon}}\gamma_\nu{i(\not{\!p_1}+\not{\!q}+m)\over{(p_1+q)^2-m^2+i\epsilon}}\gamma_\mu{i(\not{\!p_1}+m)\over{p_1^2-m^2+i\epsilon}}\Bigg]\Bigg\}.
\en
The derivation of $F^{\text{s,t}}_{\pi\gamma^*}$ is almost the same as that in (\ref{tran4},\ref{trant3}). But there is one thing needed to be clarified: $(p+q)^-$ in (\ref{tran4},\ref{trant3}) is the LF energy component of the on-shell photon, while it is the one corresponding to the off-shell photons in $F^{\text{s,t}}_{\pi\gamma^*}$. This difference will become more clear in subsection D.
\subsection{Light-Front Formalism}
Within the light-front formalism, a meson consisting of a quark $q_1$ and
an antiquark $\bar q_2$ with total momentum $p$
and spin $S$ can be written as
\begin{eqnarray}
        |M(p, S, S_z)\rangle
                &=&\int \{d^3p_1\}\{d^3p_2\} ~2(2\pi)^3 \delta^3(\tilde
                p-\tilde p_1-\tilde p_2) \non \\
                &&\sum_{\lambda_1,\lambda_2}
                \Psi^{SS_z}(\tilde p_1,\tilde p_2,\lambda_1,\lambda_2)~
                |q_1(p_1,\lambda_1) \bar q_2(p_2,\lambda_2)\rangle,
\end{eqnarray}
where $p_1$ and $p_2$ are the on-shell light-front momenta,
\begin{equation}
        \tilde p=(p^+, p_\bot)~, \quad p_\bot = (p^1, p^2)~,
                \quad p^- = {m^2+p_\bot^2\over p^+},
\end{equation}
and
\begin{eqnarray}
        &&\{d^3p\} \equiv {dp^+d^2p_\bot\over 2(2\pi)^3}, \nonumber \\
        &&|q(p_1,\lambda_1)\bar q(p_2,\lambda_2)\rangle
        = b^\dagger_{\lambda_1}(p_1)d^\dagger_{\lambda_2}(p_2)|0\rangle,\\
        &&\{b_{\lambda'}(p'),b_{\lambda}^\dagger(p)\} =
        \{d_{\lambda'}(p'),d_{\lambda}^\dagger(p)\} =
        2(2\pi)^3~\delta^3(\tilde p'-\tilde p)~\delta_{\lambda'\lambda}.
                \nonumber
\end{eqnarray}
In terms of the light-front relative momentum
variables $(x, k_\bot)$ defined by
\begin{eqnarray}
        && p^+_1=(1-x) p^+,~~~~~~~~~ \quad p^+_2=x p^+, \nonumber \\
        && p_{1\bot}=(1-x) p_\bot+k_\bot, \quad p_{2\bot}=x p_\bot-k_\bot, \label{pxk}
\end{eqnarray}
the momentum-space wave-function $\Psi^{SS_z}$
can be expressed as
\begin{equation}
        \Psi^{SS_z}(\tilde p_1,\tilde p_2,\lambda_1,\lambda_2)
                = R^{SS_z}_{\lambda_1\lambda_2}(x,k_\bot)~ \phi(x, k_\bot),
\end{equation}
where $\phi(x,k_\bot)$ describes the momentum distribution of the
constituents in the bound state, and $R^{SS_z}_{\lambda_1\lambda_2}$
constructs a state of definite spin ($S,S_z$) out of light-front
helicity ($\lambda_1,\lambda_2$) eigenstates.  Explicitly,
\begin{equation}
        R^{SS_z}_{\lambda_1 \lambda_2}(x,k_\bot)
                =\sum_{s_1,s_2} \langle \lambda_1|
                {\cal R}_M^\dagger(1-x,k_\bot, m_1)|s_1\rangle
                \langle \lambda_2|{\cal R}_M^\dagger(x,-k_\bot, m_2)
                |s_2\rangle
                \langle {1\over2}s_1;
                {1\over2}s_2|S,S_z\rangle,
\end{equation}
where $|s_i\rangle$ are the usual Pauli spinors,
and ${\cal R}_M$ is the Melosh transformation operator:
\begin{equation}
        {\cal R}_M (x,k_\bot,m_i) =
                {m_i+x M_0+i\vec \sigma\cdot\vec k_\bot \times \vec n
                \over \sqrt{(m_i+x M_0)^2 + k_\bot^2}},
\end{equation}
with $\vec n = (0,0,1)$, a unit vector in the $z$-direction, and
\begin{equation}
        M_0^2={ m^2_1+k_\bot^2\over {(1-x)}}+{m^2_2+k_\bot^2\over{x}}.
\label{M0}
\end{equation}
In practice it is more convenient to use the covariant form for
$R^{SS_z}_{\lambda_1\lambda_2}$ \cite{Jaus}:
\begin{equation}
        R^{SS_z}_{\lambda_1\lambda_2}(x,k_\bot)
                ={1\over {\sqrt{2} ~ {\widetilde M_0}}}
        ~\bar u(p_1,\lambda_1)\Gamma v(p_2,\lambda_2), \label{covariant}
\end{equation}
where ${\widetilde M_0} \equiv \sqrt{M_0^2-(m_1-m_2)^2}$, $\Gamma=\gamma_5~(\text{pseudoscalar, S=0})$, and
\begin{eqnarray}
       \sum_\lambda u(p,\lambda)\bar u(p,\lambda)&=&\not{\! p}+m,\,\,\, \bar u(p,\lambda) u(p,\lambda')=2\,m\delta_{\lambda,\lambda'},\non \\
       \sum_\lambda v(p,\lambda)\bar v(p,\lambda)&=&\not{\! p}-m,\,\,\, \bar v(p,\lambda) v(p,\lambda')=-2\,m\delta_{\lambda,\lambda'}.
\end{eqnarray}
We normalize the meson state as
\begin{equation}
        \langle M(p',S',S'_z)|M(p,S,S_z)\rangle = 2(2\pi)^3 p^+
        \delta^3(\tilde p'- \tilde p)\delta_{S'S}\delta_{S'_zS_z}~,
\label{wavenor}
\end{equation}
so that
\begin{equation}
        \int {dx\,d^2k_\bot\over 2(2\pi)^3}~|\phi(x,k_\bot)|^2 = 1.
\label{momnor}
\end{equation}

After considering the normalization condition, the momentum distribution function $\phi(x,k_\perp)$, for the pion, is related to the bound state vertex function $\Lambda_\pi$ by \cite{Jaus,Dem}
\be
{\Lambda_\pi \over{(p^2-M_0^2)}} \longrightarrow {\sqrt{x(1-x)}\over{\sqrt{2}~\widetilde {M}_0}}\phi_\pi(x,k_\perp). \label{laphi}
\en
Now, we can use the definitions of the light-front relative momentum
variables $(x, k_\bot)$ (\ref{pxk}) and the relation between the constant vertex function $\Lambda_\pi$ and the momentum distribution function $\phi(x,k_\perp)$ (\ref{laphi}) to rewrite the equations of those decay constants and form factors in the subsections A, B, and C. The results are the following. For the decay constant,
\be
f_\pi=\,2\sqrt{N_c}\int {dx\,d^2k_\perp\over 2(2\pi)^3}\,\phi_\pi(x,
k_\perp)\,{m\over\sqrt{m^2+k_\perp^2}}. \label{fp}
\en
For charge form factor in the space-like region,
\be
F^{\text{s}}_\pi (q^2) = \int {dx~d^2 k_\perp\over {2 (2\pi)^3}}\phi_{\pi'}(x,k'_\perp) \phi_\pi(x,k_\perp) {\widetilde{M}_0\over {\widetilde{M'}_0}} \Bigg(1+{x q_\perp \cdot k_\perp \over {m^2+k_\perp^2}}\Bigg),\label{FP}
\en
where $k'_\perp=k_\perp+x q_\perp$ and
\be
M_0^{'2}={ m^2+k^{'2}_\bot\over (1-x)}+{ m^2+k^{'2}_\bot\over x}.
\en
For charge form factor in the time-like region,
\be
F^{\text{t}}_\pi (q^2) &=& \int {d^2 k_\perp\over {2 (2\pi)^3}}\Bigg(\int^r_0 dx {2\Bigg[{x(1-x)\over{r}}M^2_\pi-r(m^2+k^2_\perp)\Bigg]\Lambda_{\pi'}\over{r(1-x)\Bigg[q^2-{m^2+k^2_\perp\over{x(1-x)}}\Bigg]}}{\phi_\pi(x/r,k_\perp)\over{\sqrt{2}\sqrt{m^2+k^2_\perp}}} \non \\
&&~~~+\int^1_r dx {2\Bigg[(1-r)(m^2+k^2_\perp)-{x(1-x)\over{1-r}}M^2_\pi\Bigg]\Lambda_{\pi}\over{x(1-r)\Bigg[q^2-{m^2+k^2_\perp\over{x(1-x)}}\Bigg]}}{\phi_{\pi'}((x-r)/(1-r),k_\perp)\over{\sqrt{2}\sqrt{m^2+k^2_\perp}}}\Bigg), \label{EMtime}
\en
where $r=p^+/q^+$. One can find that, for the two terms in Eq. (\ref{EMtime}), there remains one bound state vertex function which can not be substitued with the momentum distribution function shown in (\ref{laphi}). In fact, this is the nonvalence contribution arising from quark-pair creation \cite{Dem,CCH}. We illustrate this situation in Fig.1 (d), the process $q\to q\bar{q} q \to q\pi$ is represented by the empty circle. Another $\phi_\pi(x,k_\perp)$ is not applicable for the empty circle because the light-front momentum of the quark is larger than that of the daughter pion (i.e. $p^+_1\geq p^+$). This makes the task of calculating the nonvalence contribution considerably difficult.

For transition form factors in the space-like region,
\be
&&F^{\text{s}}_{\pi\gamma}(q^2) = 2{\sqrt{N_c}\over{3}}\int {dx\,d^2k_\perp\over 2(2\pi)^3}\,\phi_\pi(x,
k_\perp)\,{m\over\sqrt{m^2+k_\perp^2}} \non \\
&&~~~~~\times\Bigg[{1\over{(1-x)q^2_\perp+xM_0^2-2k_\perp\cdot q_\perp }}+{1\over{xq^2_\perp +(1-x)M^2_0+2k_\perp\cdot q_\perp}}\Bigg],\label{FMG}
\en
and
\be
&&F^{\text{s}}_{\pi\gamma^*}(q^2,q'^2) = 2{\sqrt{N_c}\over{3}}\int {dx\,d^2k_\perp\over 2(2\pi)^3}\,\phi_\pi(x,
k_\perp)\,{m\over\sqrt{m^2+k_\perp^2}} \non \\
&&~~\times\Bigg[{1\over{(1-x)(q^2_\perp-q'^2_\perp)+xM_0^2-2k_\perp\cdot q_\perp }}+{1\over{x(q^2_\perp-q'^2_\perp)+(1-x)M^2_0+2k_\perp\cdot q_\perp}}\Bigg].\label{FMGS}
\en
For transition form factors in the time-like region,
\be
F^{\text{t}}_{\pi\gamma}(q^2)&=& 2{\sqrt{N_c}\over{3}}\int{d^2k_\perp\over 2(2\pi)^3}{m\over\sqrt{m^2+k_\perp^2}} \non \\
&&\times \Bigg(\int^{1-y}_0 dx \,\,\phi_\pi(x,k_\perp)\,{x\over{(1-y)(m^2+k^2_\perp)}}\non \\
&&~~~~+\int^1_y dx\,\,\phi_\pi(x,k_\perp)\, {1-x\over{(1-y)(m^2+k^2_\perp)}} \non \\
&&~~~+\int^1_{1-y} dx\,\,\phi_\pi(x,
k_\perp)\,{1-x\over{y(m^2+k^2_\perp)-(1-x)(x-1+y)M^2_\pi}} \non \\
&&~~~+\int^y_0 dx\,\,\phi_\pi(x,
k_\perp)\,{x\over{y(m^2+k^2_\perp)-x(y-x)M^2_\pi}}\Bigg)\label{Ftime1}
\en
and
\be
F^{\text{t}}_{\pi\gamma^*}(q^2,q'^2)&=& 2{\sqrt{N_c}\over{3}}\int{d^2k_\perp\over 2(2\pi)^3}{m\over\sqrt{m^2+k_\perp^2}}\non \\
&&\times\Bigg(\int^{1-y}_0 dx \,\,\phi_\pi(x,k_\perp)\,{x(1-y)\over{(1-y)^2(m^2+k^2_\perp)-x(1-y-x)q'^2}}\non \\
&&~~~~+\int^1_y dx\,\,\phi_\pi(x,k_\perp)\, {(1-x)(1-y)\over{(1-y)^2(m^2+k^2_\perp)-(1-x)(x-y)q'^2}} \non \\
&&~~~+\int^1_{1-y} dx\,\,\phi_\pi(x,
k_\perp)\,{(1-x)y\over{y^2(m^2+k^2_\perp)-(1-x)(x-1+y)q^2}} \non \\
&&~~~+\int^y_0 dx\,\,\phi_\pi(x,
k_\perp)\,{xy\over{y^2(m^2+k^2_\perp)-x(y-x)q^2}}\Bigg), \label{Ftime2}
\en
where $y\equiv q^+/p^+$ and
\be
y={M^2_\pi+q^2-q'^2-\sqrt{(M^2_\pi+q^2-q'^2)^2-4M^2_\pi \,q^2}\over{2\,M^2_\pi}}.
\en
We can easily check that $F^{\text{s}}_{\pi\gamma}(0)=F^{\text{t}}_{\pi\gamma}(0)\equiv F_{\pi\gamma}(0)$.
\section{Numerical Results and Discussions}
We now compare our results for form factors with the experimental data. Before doing this, we need to determine the parameters appearing in the wave function $\phi_\pi(x,k_\perp)$. Of course, it is assumed that this wave function is universal and processes-independent.

One wave function that has been often used in the literature for mesons is the Gaussian-type,
\begin{equation}
    \phi(x,k_\perp)_{\rm G}={\cal N} \sqrt{{dk_z\over dx}}
        ~{\rm exp}\left(-{\vec k^2\over 2\omega^2}\right),
        \label{gauss}
\end{equation}
where ${\cal N}=4(\pi/\omega^2)^{3/4}$ and $k_z$ is of the internal momentum
$\vec k=(\vec{k}_\bot, k_z)$, defined through
\begin{equation}
1-x = {e_1-k_z\over e_1 + e_2}, \qquad
x = {e_2+k_z \over e_1 + e_2},
\end{equation}
with $e_i = \sqrt{m_i^2 + \vec k^2}$. We then have
\be
M_0=e_1 + e_2,~~~~k_z = \,{xM_0\over 2}-{m_2^2+k_\perp^2 \over 2 xM_0},
\label{kz}
\en
and
\begin{equation}
        {{dk_z\over dx}} = \,{e_1 e_2\over x(1-x)M_0},
\end{equation}
which is the Jacobian of transformation from $(x, k_\bot)$ to $\vec k$. This wave function has been also used in many other studies
of hadronic transitions. Besides this wave function which has the exponential form, we also consider the power-law type one
\be
\phi(x,k_\perp)_N=\,{\cal N}\Bigg({\omega^2\over{M_0^2+\omega^2}}\Bigg)^{n}.  \label{hwcw1}
\en
where $n$ is another parameter.

For the pion case, we assume that the constituent mass of the $u$ and $d$ quarks is the same, i.e., $m_u=m_d\equiv m$. Thus, there are two parameters $m$ and $\omega$ in the Gaussian-type wave function and an additional parameter $n$ in the power-law type one. We will use three conditions to fix these parameters. The first one is the experimental value of the decay constant $f_\pi=92.4\text{MeV}$\cite{PDG00} and (\ref{fp}). The second one is the electromagnetic radius of the charged pion
\be
\langle r^2\rangle_\pi\simeq-6{\partial F_\pi(q^2)\over{\partial q^2}}\Bigg |_{q^2=0}.
\en
where $\langle r^2\rangle^{\text{exp}}_\pi=0.439\pm0.03~fm^2$ \cite{Dally}. The third one is the decay width $\Gamma\,(\pi^0\to\gamma\gamma)$  \cite{PDG00}. It is well known that $F_{\pi\gamma}(0)$ can be determined from this decay width via \cite{CLEOFpgs}
\be
|F_{\pi\gamma}(0)|^2={1\over{(4\pi\alpha)^2}}{64\pi\,\Gamma(\pi^0\to\gamma\gamma)\over{M_\pi^3}}. \label{F0exp}
\en
Thus leads to $F^{\text{exp}}_{\pi\gamma}(0)=0.27\pm0.01~\text{GeV}^{-1}$. Following from Eq. (\ref{FMG}) we have the theoretical expression for $F_{\pi\gamma}(0)$
\be
F_{\pi\gamma}(0) = 2{\sqrt{N_c}\over{3}}\int {dx\,d^2k_\perp\over 2(2\pi)^3}\,\phi_\pi(x,
k_\perp)\,{m\over\sqrt{m^2+k_\perp^2}} \Bigg[{1\over{x(1-x)M_0^2}}\Bigg],\label{FMG0}
\en
Using these constraints, we can uniquely determine all the parameters in $\phi_\pi (x,k_\perp)$ (\ref{hwcw1}). Here we list the fitted parameters of these two wave functions $\phi_G$ and $\phi_N$ in Table 1. We show the $x$-dependent and $k^2_\perp$-dependent behaviors of these two types of wave functions in Fig.2 and Fig.3, respectively. Note that the location of maxima is quite different for $\phi_G$ and $\phi_N$ as shown in Fig.2 and that the transverse momentum suppression of the exponential forms $\phi_G$ is more stronger than that of $\phi_N$ (see Fig.3).

\vskip 0.5cm
\begin{center}
\begin{tabular}{|c||c|c|c|c|c|} \hline
~wave function~ & $m_q$(GeV) & $\omega_\pi$(GeV) & $f_\pi$(MeV) & $\langle r^2\rangle_\pi~(fm^2)$ & $F_{\pi\gamma}(0)~(\mbox{GeV}^{-1})$ \\ \hline
$\phi_G$ & 0.243 & 0.328 & 92.4 & 0.434 & $0.231$ \\ \hline
$\phi_N(n=1.7)$ & 0.192 & 0.957 & 92.4 & 0.434 & $0.272$ \\ \hline
\end{tabular}
\end{center}
\begin{center}
{\small Table I. Parameters $m_q$ and $\omega_\pi$ in wave functions $\phi_G$ and $\phi_N$.}
\end{center}

From Table I, we find that the value $F_{\pi\gamma}(0)=0.27~\mbox{GeV}^{-1}$ cannot be reached by adjusting the parameters $m_q$ and $\omega_\pi$ in the wave function $\phi_G$. It is well known that $\pi^0 \to \gamma\gamma$ is not only the dominant decay mode ($98.8\%$), but also relates to all the observed decay modes of the neutral pion (see below). Therefore the Gaussian wave function, in spite of being used in the literature for mesons, is not suitable to describe the decay processes of the neutral pion at least. Thus we will use the power-law type wave function $\phi (x,k_\perp)_N$ to calculate the form factors $F_\pi(Q^2)$ and $F_{\pi\gamma}(Q^2)$. The form factors $F_\pi(q^2)$ and $F^{\text{s}}_{\pi\gamma}(q^2)$, can be calculated by Eq.(\ref{FP}) and Eq.(\ref{FMG}), and the results are plotted in Fig.4, and Fig.5, respectively. They are in agreement with the experimental data. It appears that LFQM is valid up to the scale of order $q^2\sim -8~\mbox{GeV}^2$, while the constituent quark model is only applicable to the low-energy region. In the other approaches, \cite{CJ2} used the Gaussian type wave fnction and the axial anomaly plus the PCAC relations to calculate the space-like form factors $F_\pi (q^2)$ and $F^{\text{s}}_{\pi\gamma} (q^2)$, their results are in agreement with data. The pQCD approach has also been used to calculated the form factors $F_\pi(q^2)$ \cite{MNP} in $q^2\leq -4~\text{GeV}^2$ region and $F^{\text{s}}_{\pi\gamma}(q^2)$ \cite{KR,MR} in $-8~\text{GeV}^2\le q^2 \le -0.5 ~\text{GeV}^2$ region. In those regions, their results fit the data well. 

It is worth while to mention that $\pi^0 \to \gamma\gamma$ amplitude comes entirely from the anomaly in the soft-pion limit \cite{Adler}. The anomaly approach begins with \cite{CL} this amplitude, and in terms of modified partially conserved axial-vector current (PCAC) hypothesis, it relates to a relevant axial-vector amplitude. In the soft-pion limit, the $\pi^0 \to \gamma\gamma$ amplitude is proportional to an anomaly term. The major difference between this approach and ours is that the former one doesn't assume an internal structure for pion, that is, the pion couples to the quarks simply via the vertex $\gamma_5$. The pion decay constant in the anomaly approach is regarded as an input value, in contrast with this paper, it could be theoretically calculated from the integral of a quark momentum distribution function (\ref{fp}).

Now we use the wave function $\phi_\pi(x,k_\perp)$ (\ref{hwcw1}) to calculate the transition form factors in time-like region $0\leq q^2\leq 0.02~\mbox{GeV}^2$. The equation (\ref{Ftime1}) can be used to get $F^{\text{t}}_{\pi\gamma}(q^2)$ with the results shown in Fig.6. This form factor is related to the differential decay rate of $\pi^0 \to \gamma\,e^+e^-$ by
\be
{d\,\Gamma(\pi^0 \to \gamma\,e^+e^-)\over{\Gamma(\pi^0 \to \gamma\gamma)\,dq^2}}={2\over{q^2}}\left({\alpha\over{3\pi}}\right)\left|{F^{\text{t}}_{\pi\gamma}(q^2)\over{F_{\pi\gamma}(0)}}\right|^2\,\lambda^{3/2} \left(1,{q^2\over{M^2_{\pi^0}}},0\right)\,G_e(q^2),
\en
where
\be
\lambda (a,b,c)=a^2+b^2+c^2-2(ab+bc+ca),
\en
and
\be
G_e(q^2)=\left(1-{4\,M^2_e\over{q^2}}\right)^{1/2}\left(1+{2\,M^2_e\over{q^2}}\right).
\en
We take the integration of $q^2$ and obtain the branching ratio
\be
{\cal B}_{\gamma e^+e^-}={\Gamma(\pi^0\to\gamma e^+e^-)\over{\Gamma(\pi^0\to\gamma \gamma)}}=1.244\times 10^{-2}
\en
This value agrees well with the data $(1.25\pm 0.04\pm 0.01)\times 10^{-2}$ \cite{lampf} and the weighted average of the two experimental results $(1.213\pm 0.033)\times 10^{-2}$ \cite{PDG00}. In the approach of pure QED (neglecting strong-interaction effects), ${\cal B}_{\gamma e^+e^-}$ has also been calculated and is found to be $1.185 \times 10^{-2}$ \cite{KW}. The approach of VMD \cite{Babu} predicts the mass of the vector meson to be $M_V= 0.821\pm0.015$GeV and also sets $F^{\text{t}}_{\pi\gamma}(q^2) \to F_{\pi\gamma}(0)$ approximately.

Another decay mode related to $F^{\text{t}}_{\pi\gamma}(q^2)$ is $\pi^0 \to\gamma+\text{positronium}$. This process occurs in the Dalitz decay $\pi^0\to e^+e^-\gamma$ below slightly the lowest possible invariant mass of the lepton pair since the Coulomb binding energy is negative. Owing to $4m^2_e\ll M^2_\pi$, the ratio $F^{\text{t}}_{\pi\gamma}(4m^2_e)/F^{\text{t}}_{\pi\gamma}(0)\approx 1$. Therefore the QCD correction to the branching ratio is very small and the branching ratio ${\cal B}_{\gamma+\text{positronium}}$ becomes \cite{Nemenov}
\be
{\Gamma(\pi^0\to\gamma+\text{positronium})\over{\Gamma(\pi^0\to\gamma\gamma)}}=32\pi\alpha\Bigg(1-{4m^2_e\over{M^2_\pi}}\Bigg)\sum^\infty_{n=1} {|\Psi_{nLj}(0)|^2\over{4m^2_e}}\approx 0.6\alpha^4\approx 1.7\times 10^{-9}
\en
where $\Psi_{nLj}$ is the nonrelativistic bound state. The experimental value
${\cal B}^{\text{exp}}_{\gamma+\text{positronium}}=(1.84\pm 0.29)\times 10^{-9}$ \cite{PDG00}.

Next, the equation of (\ref{Ftime2}) can be used to calculate $F^{\text{t}}_{\pi\gamma^*}(q^2,q'^2)$. This form factor is related to the differential decay rate of $\pi^0 \to e^+e^+e^-e^-$ by
\be
{d\,\Gamma(\pi^0 \to e^+e^+e^-e^-)\over{\Gamma(\pi^0 \to \gamma\gamma)\,dq^2\,dq'^2}}&=&{1\over{q^2q'^2}}\left({\alpha\over{3\pi}}\right)^2\left|{F^{\text{t}}_{\pi\gamma^*}(q^2,q'^2)\over{F_{\pi\gamma}(0)}}\right|^2\,\non \\
&&\times\lambda^{3/2} \left(1,{q^2\over{M^2_{\pi^0}}},{q'^2\over{M^2_{\pi^0}}}\right)\,G_e(q^2)\,G_e(q'^2).
\en
After the integrations over $q^2$ and $q'^2$ we obtain the branching ratio
\be
{\cal B}_{e^+e^+e^-e^-}={\Gamma(\pi^0\to e^+e^+e^-e^-)\over{\Gamma(\pi^0\to\gamma \gamma)}}=3.00\times 10^{-5},
\en
which agrees well with the data $(3.18\pm 0.30)\times 10^{-5}$ \cite{sam}. The prediction of ${\cal B}_{e^+e^+e^-e^-}$ in pure QED \cite{KW} is $3.46 \times 10^{-5}$.

The tranistion form factor $F_{\pi\gamma^*}(q^2,q'^2)$ is also related to another decay rate $\Gamma(\pi^0 \to e^+e^-)$ (for the lowest order in QED) by \cite{Berg}
\be
{\cal B}_{e^+e^-}={\Gamma(\pi^0\to e^+e^-)\over{\Gamma(\pi^0\to\gamma \gamma)}}= 2\left({\alpha\over{\pi}}\right)^2 \left(1-{4\,M^2_e\over{M^2_{\pi_0}}}\right)^{1/2}\left({M_e\over{M_{\pi^0}}}\right)^2\,|{\cal R}|^2
\en
where
\be
{\cal R}(p^2)={2i\over{\pi^2 M_{\pi^0}^2}}\int d^4 q\,{[M^2_{\pi^0}q^2-(p\cdot q)^2]\over{q^2\,(p-q)^2\,[(q-p_e)^2-M^2_e]}}\, {F_{\pi\gamma^*}(q^2,(p-q)^2)\over{F_{\pi\gamma}(0)}} \label{Rloop}
\en
It is well known that the unitarity bound for ${\cal B}(\pi^0 \to e^+e^-)$ comes from the on-shell $\gamma\gamma$ intermediate state. This state generates the model-independent imaginary part of ${\cal R}$ \cite{Drell}
\be
\text{Im}\,\,{\cal R}(M^2_\pi)=\left({\pi\over{2\,\beta_\pi}}\right)\,\text{ln}\left[{1-\beta_\pi}\over{1+\beta_\pi}\right], \label{imaginary}
\en
where $\beta^2_\pi\equiv 1-4M^2_e/M^2_{\pi^0}$. Eq.(\ref{imaginary}) gives rise to the unitary limit
\be
{\cal B}_{e^+e^-}\geq{\cal B}^{\text{unit}}_{e^+e^-}=4.75\times 10^{-8}.
\en
As for the real part of ${\cal R}$, we find, in Eq. (\ref{Rloop}), the integration is too difficult to do because that $q^2$ and $(p-q)^2$ don't have to larger than zero simultaneously. Thus Re ${\cal R}$ can be estimated by applying Eq. (\ref{FMGS}) in the soft pion limit, $p\to 0$. Then one obtains
\be
\text{Re}\,\,{\cal R}(p^2=0)\simeq -23.33.
\en
In general, an once-subtracted disperson relation can be written for $\text{Re}\,\,{\cal R}$ \cite{BMAB}
\be
\text{Re}\,\,{\cal R}(p^2)=\text{Re}\,\,{\cal R}(0)+{p^2\over{\pi}}\int^\infty_0 dp'^2{\text{Im}\,\,{\cal R}(p'^2)\over{(p'^2-p^2)p'^2}}.
\en
We find that $\text{Re}\,\,{\cal R}(M^2_\pi)=8.93$ and ${\cal B}_{e^+e^-}=5.98\times 10^{-8}$. It is in agreement with the experimental data $ (6.3\pm 0.5)\times 10^{-8}$. The method of VMD \cite{ABBM} yields the mass of the vector meson $M_V\simeq 0.77$GeV and also obtains a value $6.41\times 10^{-8}$ which is consistent with experiment.
\section{Conclusion}
In this work, the charge and transition form factors of the pion have been considered in LFQM. The parameters appearing in the light-front wave function are deterimined by experimental data of the decay constant, electromagnetic radius of the charged pion, and two-photon decay width of the neutral pion. Then we used these formulae and wave functions to calculate the charge and transition form factors of the pion and the branching ratios of all the observed decay modes of the neutral pion. The obtained results are all consistent with data. Table II listed the comparsion with experiments and other approaches.

\vskip 0.5cm
\begin{center}
\begin{tabular}{|c||c|c|c|c|} \hline
~Br~ & exp't value \cite{PDG00} & this work & VMD & pure QED  \\ \hline
${\cal B}_{\gamma e^+e^-}$ & $(1.213\pm 0.033)\times 10^{-2}$ & $1.244\times 10^{-2}$ & $1.185\times 10^{-2}$\cite{Babu} & $1.185\times 10^{-2}$\cite{KW}  \\ \hline
${\cal B}_{\gamma + \text{positronium}}$ & $(1.84\pm0.29)\times 10^{-9}$ & $1.7\times 10^{-9}$ & - & $1.7\times 10^{-9}$ \\ \hline
${\cal B}_{e^+ e^+ e^- e^-}$ & $(3.18\pm0.30)\times 10^{-5}$ & $3.00\times 10^{-5}$ & - & $3.46\times 10^{-5}$ \cite{KW}\\ \hline
${\cal B}_{e^+e^-}$ & $(6.3\pm 0.5)\times 10^{-8}$ & $5.98\times 10^{-8}$ & $6.41\times 10^{-8}$ \cite{ABBM} & $\geq 4.75\times 10^{-8}$  \\ \hline
\end{tabular}
\end{center}
\begin{center}
{\small Table II. Branching ratios for the neutral pion in this work and in some other models.}
\end{center}

The time-like region of the charge form factor was considered to be applied to the decay process $\gamma^* \to \pi\pi$, the calculation was not finished because that we do not have a reliable estimate of the pair-creation effect. Contrary to the case of charge form factor, the time-like region of the transition form factors were studied completely. The major reason for this difference is that the pair-creation effect will exist when there are hadrons in the final state. Concerning the model-dependent part, although the choice of the light-front wave function is arbitrary, it is assumed to be processes-independent. Therefore it is desirable if more data in the low-energy region, where the constituent quark model is more applicable, are fitted. This is the reason why we use the power-law type wave function. The major differences between the Gaussian type and power-law one are, first, the locations of maxima is quite different for $x$; second, the transverse momentum suppression of the exponential function is stronger than that of the power-law ($n=1.7$) function. We shall investigate the adaptability of this wave function and the treatment of Z graph for other hadrons and other processes in the near future.
\vskip 2.0cm
\centerline {\bf ACKNOWLEDGMENTS}

  This work was supported in part by the National Science Council of ROC under Contract No. NSC89-2112-M-007-054.

\newcommand{\bi}{\bibitem}
%

\newpage
\parindent=0 cm
\centerline{\bf FIGURE CAPTIONS}
\vskip 0.5 true cm

{\bf Fig. 1 } Diagrams for (a) a charged pion decay, (b) the scattering of one virtual photon and one pion, (c) one time-like photon and two pion coupling, (d) the illustration of the nonvalence contribution in (c), (e) the $\gamma^* \pi^0 \to \gamma $ vertex, and (f) the $\gamma^* \pi^0 \to \gamma^* $ vertex.
\vskip 0.25 true cm

{\bf Fig. 2 } The $x$-dependent behavior of $\phi_G$ and $\phi_N~(n=1.7)$ at $k_\perp=0$.
\vskip 0.25 true cm

{\bf Fig. 3 } The $k^2_\perp$-dependent behavior of $\phi_G$ and $\phi_N~(n=1.7)$ at $x=0.19$, where $\phi_G(x,0)\simeq \phi_N(x,0)$ at this location.
\vskip 0.25 true cm

{\bf Fig. 4 } The charge form factor of the pion in small and large momentum transfers. Data are taken from \cite{Amen} for small momentum transfers and \cite{Bebek} (empty circles), \cite{Volmer} (filled triangles), and \cite{Brauel} (filled square) for large momentum transfers.
\vskip 0.25 true cm

{\bf Fig. 5 } The one off-shell photon transition form factor of the pion. The dotted line is the limiting behavior $2 f_\pi$ ($0.185$ GeV) predicted by pQCD. Data are taken from \cite{CLEOFpgs}.
\vskip 0.25 true cm

{\bf Fig. 6 } The $y$-dependent behavior of the function $f(y)$, where $f(y)\equiv |F^{\text{t}}_{\pi\gamma}(q^2)/F_{\pi\gamma}(0)|^2$ and $y=q^2/M^2_\pi$. The dotted line is the assumption of pure QED: $f(y)=1$.
\vskip 0.25 true cm

\newpage

\begin{figure}[h]
\hskip 5.7cm
\hbox{\epsfxsize=11.2cm
      \epsfysize=14cm
      \epsffile{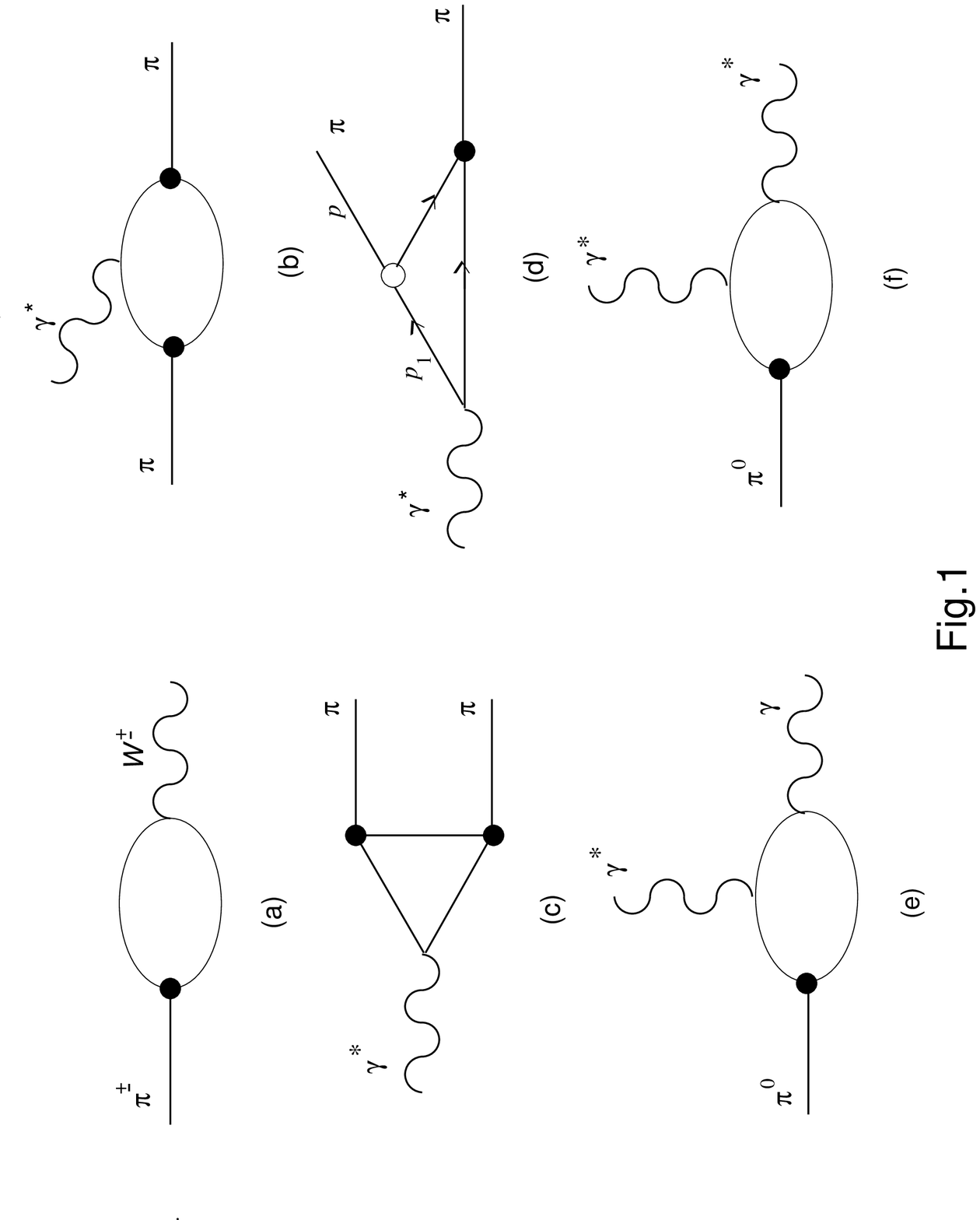}}
\end{figure}
\newpage

\begin{figure}[h]
\hskip 4cm
\hbox{\epsfxsize=16cm
      \epsfysize=20cm
      \epsffile{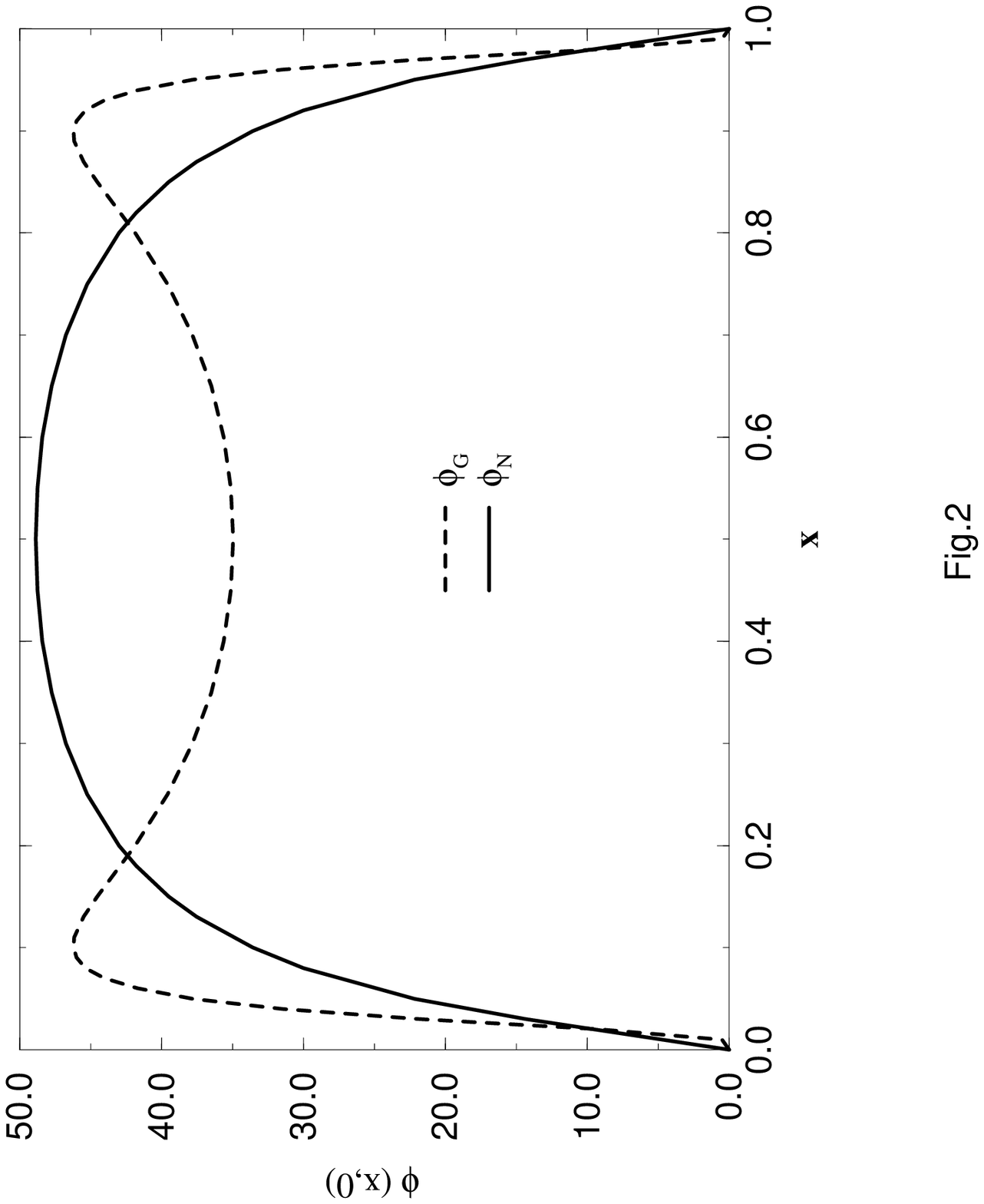}}
\end{figure}
\newpage

\begin{figure}[h]
\hskip 4cm
\hbox{\epsfxsize=16cm
      \epsfysize=20cm
      \epsffile{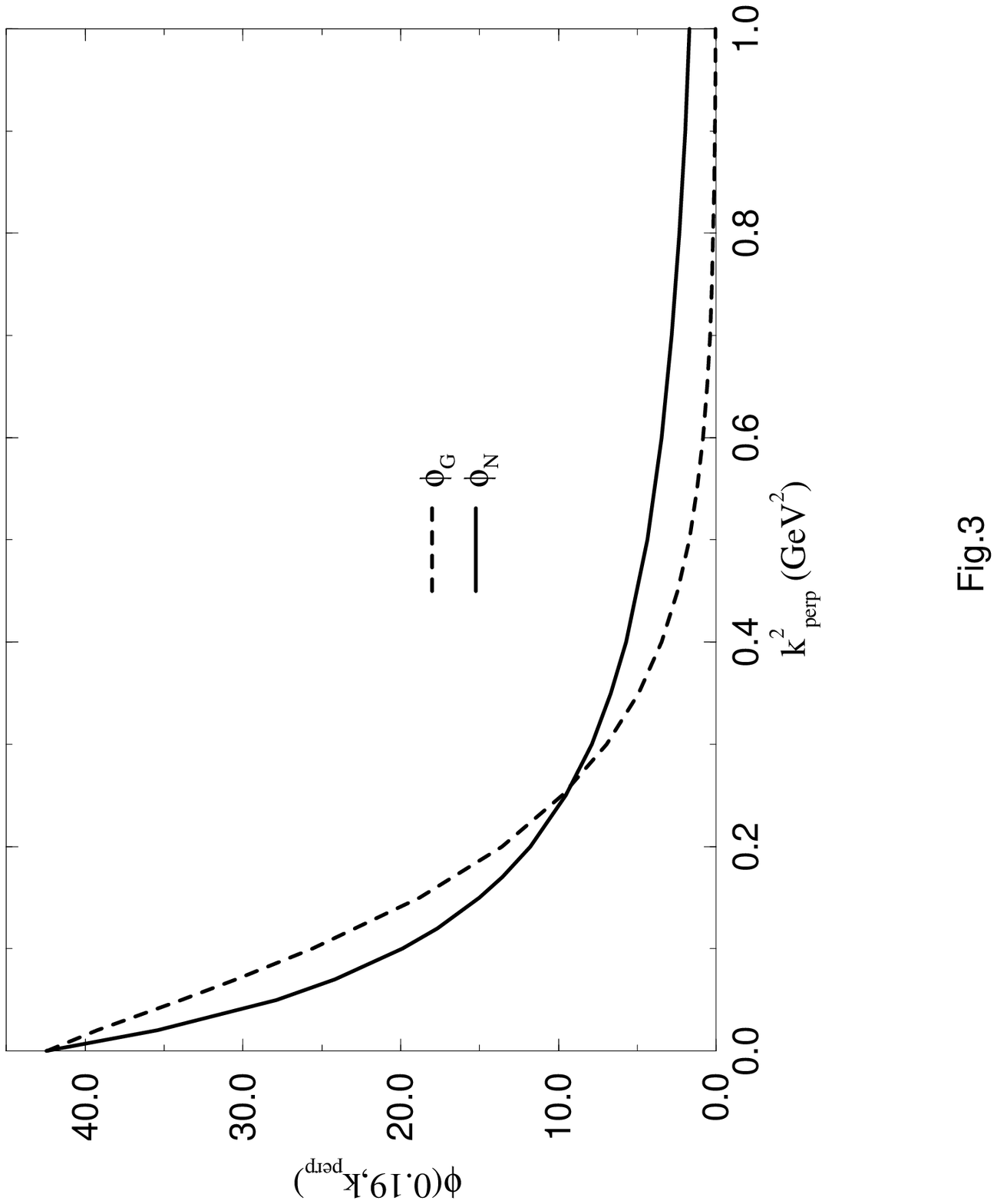}}
\end{figure}
\newpage

\begin{figure}[h]
\hskip 4cm
\hbox{\epsfxsize=16cm
      \epsfysize=20cm
      \epsffile{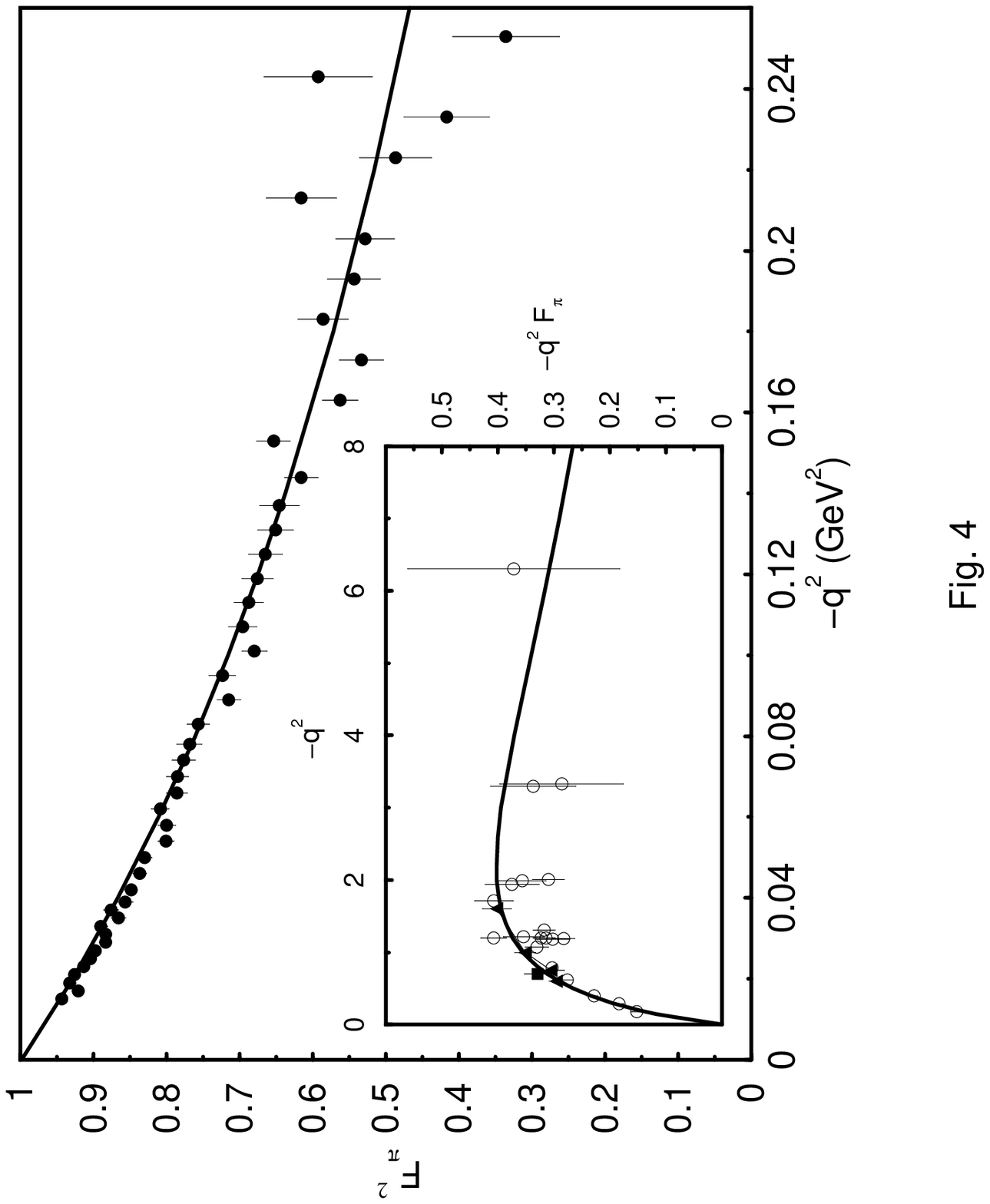}}
\end{figure}
\newpage

\begin{figure}[h]
\hskip 5.7cm
\hbox{\epsfxsize=16cm
      \epsfysize=20cm
      \epsffile{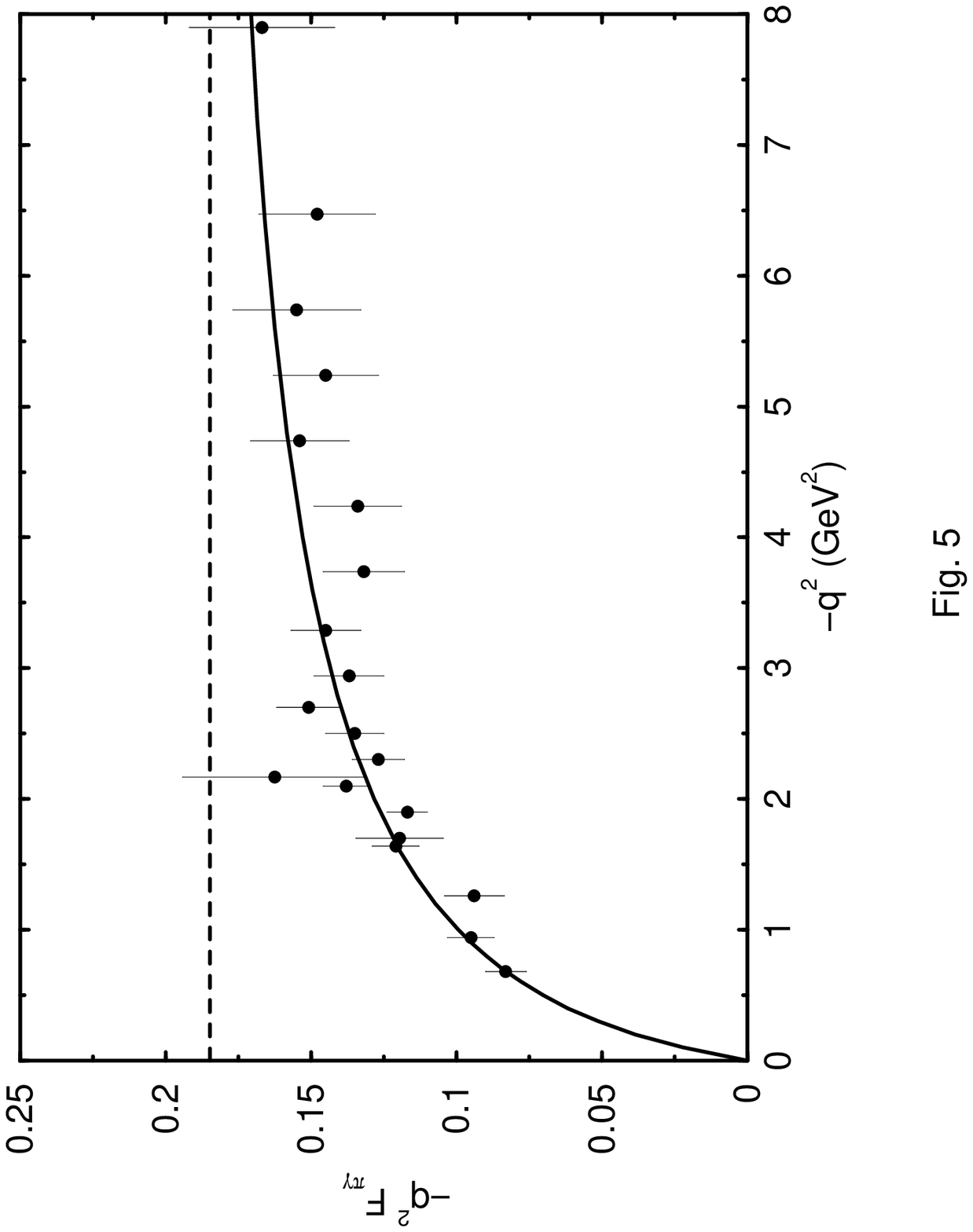}}
\end{figure}
\newpage

\begin{figure}[h]
\hskip 4cm
\hbox{\epsfxsize=16cm
      \epsfysize=20cm
      \epsffile{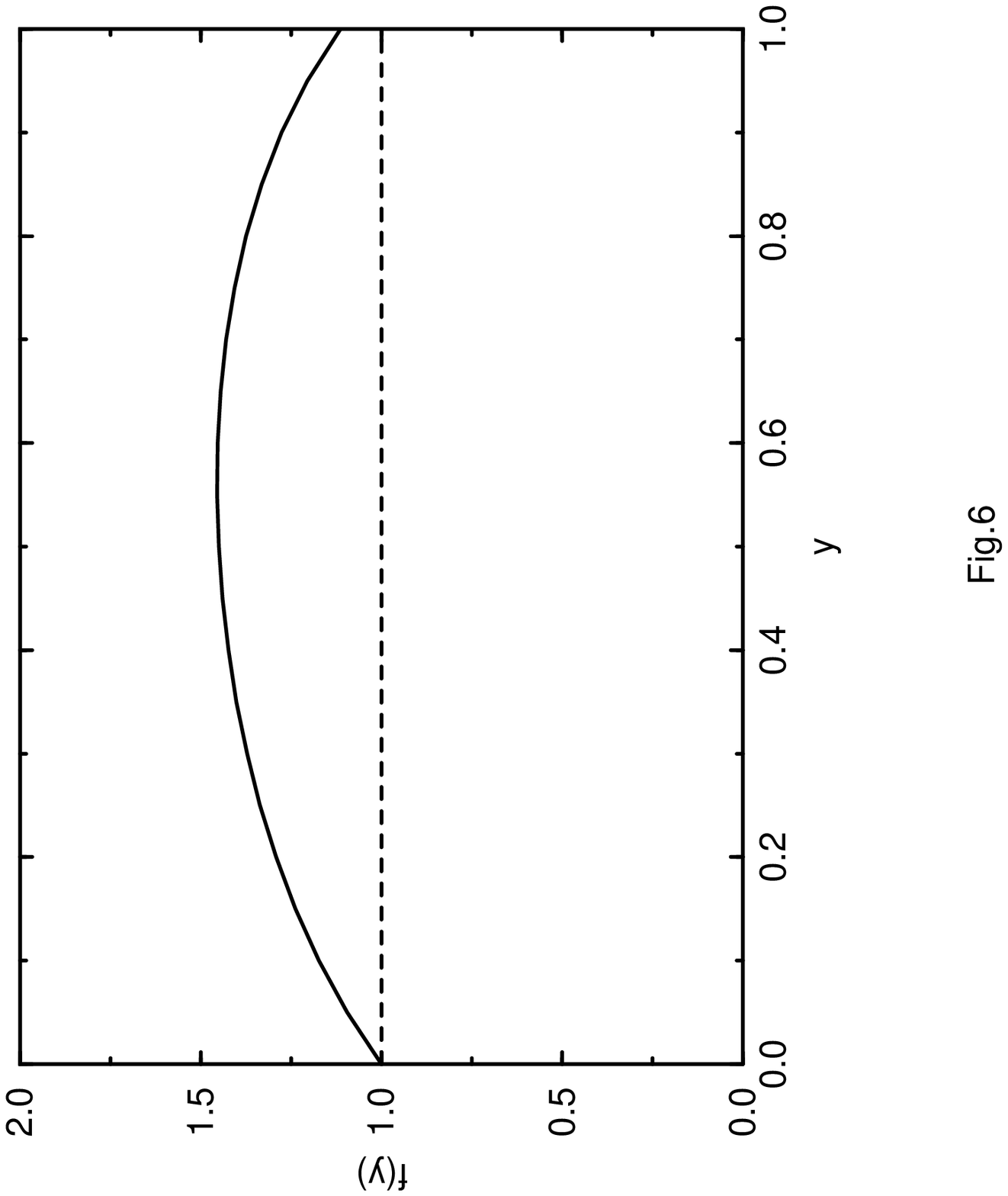}}
\end{figure}
\newpage

\end{document}